\documentclass[aps,superscriptaddress,eqsecnum,nofootinbib,preprintnumbers]{revtex4-2}
\usepackage{amsmath}
\usepackage{amsfonts}
\usepackage{amssymb}
\usepackage{commath}
\usepackage{graphicx}
\usepackage{dcolumn}
\usepackage{tabularx,ragged2e,booktabs}
\newcolumntype{C}{>{\Centering\arraybackslash}X}
\usepackage{multirow} 
\usepackage{natbib}
\usepackage{subfigure}
\usepackage{dashbox}

\usepackage{caption}
\usepackage{hyperref}
\usepackage{array}
\numberwithin{equation}{section}
\renewcommand{\thesection}{\arabic{section}}

\begin{document}
\title{Viable Cosmological Solutions from Hybrid Potentials}
\author{Koralia Tzanni}
\email{tzanni@aegean.gr}
\affiliation{Department of Environment, University of the Aegean, University Hill, Mytilene 81100, Greece}
\author{John Miritzis}
\email{imyr@aegean.gr}
\affiliation{Department of Marine Sciences, University of the Aegean, University Hill, Mytilene 81100, Greece}
\date{\today}



\begin{abstract}
We study flat Friedmann-Lema\^{\i}tre-Robertson-Walker (FLRW) models with a
perfect fluid matter source and a scalar field minimally coupled to matter
with power-law-exponential \textquotedblleft hybrid\textquotedblright potential. Using expansion-normalised variables, we formulate the field equations as a constrained three-dimensional dynamical system and determine its equilibrium structure. We show that viable cosmological histories, consisting of a transient matter or radiation era followed by late-time accelerated expansion, arise in restricted regions of parameter space. A central result is that the physically relevant trajectories are confined to an invariant plane, which contains both the transient matter point $\mathcal{B}$ and the accelerated point $\mathcal{C}$. We further show, by centre-manifold analysis, that the accelerated point $\mathcal{C}$ is not a global attractor: it attracts trajectories with $\phi>0$ and repels those with $\phi<0$. For dust, a standard matter era requires vanishing coupling of the scalar field to matter, while for radiation the interaction term vanishes identically. Finally, we discuss the issue that the qualitative cosmological
dynamics may be independent of the precise functional form of the scalar-field potential.
\end{abstract}

\maketitle

\section{Introduction}

The expansion history of the Universe is marked by two distinct phases of accelerated expansion: an early inflationary epoch \cite{Bassett_2006, ellis2026inflation2025} and the present late-time acceleration \cite{Frieman_2008,Weinberg_2013}. Scalar fields provide a unified framework for modeling both eras \cite{leon21, Ji_2024}. While the observed acceleration typically requires scalar fields with non-negative potentials to effectively act as a cosmological term, potentials that attain negative values cannot be excluded from realistic cosmological scenarios--see, e.g., \cite{ffkl,tzmi, gmt1, gmt2}  for discussion and motivation. 

\begin{figure}[!h]
\centering
\includegraphics[width=2.5in]{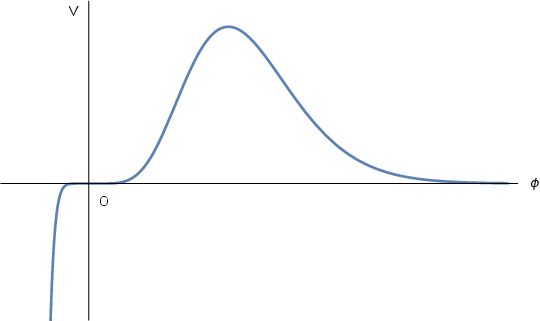} %
\includegraphics[width=2.5in]{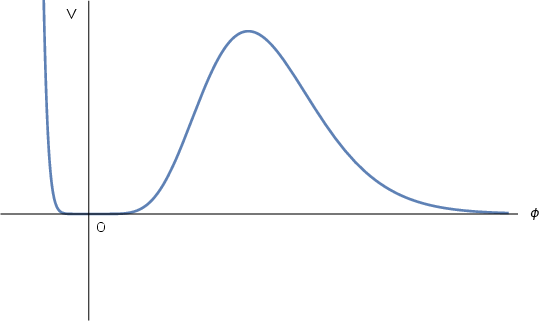}
\caption{The typical form of the hybrid potential $V\left( \protect\phi%
\right) =V_{0}\protect\phi^{n}e^{-k\protect\phi}, n\geq2$, plotted for odd
(left) and even (right) values.}
\label{fig:all_pote}
\end{figure}

In this work, we focus on a class of models containing a scalar-field coupled to ordinary matter with a constant coupling parameter and a potential of the form
\begin{equation}
V(\phi) = V_{0}\,\phi^{n}e^{-k\phi}, \label{pote}
\end{equation}
where $V_{0}>0$ and $k>0$ and $n \in \mathbb{N}$. The parity of $n$ qualitatively affects the behaviour of the potential. For even $n$ the potential remains non-negative, whereas for odd values of $n$ it can cross zero and become negative, as illustrated in Fig. \ref{fig:all_pote} for $n\geq2$. The case $n=1$ is simpler, since the potential does not exhibit the inflection--type behaviour at $\phi=0$ that occurs for odd values $n>2$.

Power--law, and in particular inverse power--law, scalar-field potentials of the form $V(\phi)\propto \phi^{n}$ have been widely studied in quintessence cosmology; see, for example, \cite{pera88,liddle99,steinhardt99,WangSteinhardt1998,BartoloPietroni1999,UrenaLopezMatos2000}. 
Exponential potentials, on the other hand, arise naturally in cosmological model building and have been extensively discussed in the contexts of both early-universe inflation and late-time acceleration; see for example the review papers \cite{Copeland2006,Tsujikawa2013}. 
Their qualitative phase-space structure was systematically analysed within the dynamical-systems framework in \cite{clw}.

Hybrid exponential--power--law potentials of the form \eqref{pote} arise in effective scalar-field cosmology and can also emerge in scalar--tensor or modified-gravity theories after suitable conformal transformations; see, for example, \cite{paba,Copeland2006,Tsujikawa2013,clw,fujimae03,faraoni04}.
For recent dynamical-systems studies of the cosmological background evolution characterized by different scalar field models see for example, \cite{tamanini14,skugoreva19,Alho22,ashi25}. For a study of quintessence models with potentials given by the product of polynomial and exponential terms that exhibit realistic accelerating solutions, see \cite{skordis}. The same potential was used to describe the dilaton field in holographic models \cite{kiritsis25}. 
Such behaviour is also consistent with the interpretation of cosmic acceleration as a bifurcation phenomenon in the underlying dynamical system \cite{cotsakis25}. 
More broadly, these ideas fit naturally within the mathematical-cosmology framework surveyed in \cite{cotsyefr22}, where attractors, singularities, and stability methods are discussed in a unified perspective.

The main contribution of the present work is threefold. First, we analyse a coupled scalar-field cosmology with hybrid power-law-exponential potential in a dynamical-systems framework that accommodates both non-negative and sign-changing behaviour of the potential. Second, we show that the viable cosmological trajectories live on an invariant plane, which contains the physically relevant matter and accelerated critical points. We note here that, throughout this work, by a viable cosmological history we mean a trajectory that passes near a transient matter- or radiation-dominated saddle with the appropriate behaviour of the scale factor and subsequently approaches a late-time accelerated state. Third, we demonstrate, using centre-manifold reduction, that the accelerated point $\mathcal{C}$ is only conditionally stable: it attracts trajectories with $\phi>0$, but repels those with $\phi<0$. This provides a geometrically transparent description of when the model yields a realistic cosmological history. 

The plan of the paper is as follows. In Section \ref{sec:ds} we formulate the field equations for flat FLRW models and rewrite them as a constrained three-dimensional dynamical system using expansion-normalised variables. In Section \ref{sec:viable} we classify the equilibrium points and determine the conditions under which a transient matter-dominated or radiation-dominated epoch is followed by accelerated expansion. We summarise the main results and raise the issue that the precise functional form of the potential has little effect on the
qualitative cosmological dynamics. Appendix \ref{sec:centre} contains the centre-manifold analysis near the accelerated critical point.

\section{Scalar field coupled to matter}\label{sec:ds}

We assume that ordinary matter is described by a perfect fluid with equation
of state 
\[
p=(\gamma-1)\rho,
\]
where $0<\gamma<2$. The limiting cases of $\gamma=0$ and $\gamma=2$ will not
be considered here. For spatially flat, homogeneous, and isotropic
spacetimes, the Einstein equations reduce to: the Friedmann constraint, 
\begin{equation}
3H^{2}=\rho+\frac{1}{2}\dot{\phi}^{2}+V\left( \phi\right) ;  \label{frie}
\end{equation}
the Raychaudhuri equation,
\begin{equation}
\dot{H}=-\frac{1}{2}\dot{\phi}^{2}-\frac{\gamma}{2}\rho;  \label{ray}
\end{equation}
the scalar field equation,
\begin{equation}
\ddot{\phi}+3H\dot{\phi}+\frac{dV\left( \phi\right)}{d\phi} =\frac{4-3\gamma%
}{2}Q\rho;  \label{ems}
\end{equation}
and the energy conservation equation,
\begin{equation}
\dot{\rho}+3\gamma\rho H=-\frac{4-3\gamma}{2}Q\rho\dot{\phi}.  \label{conss}
\end{equation}
Here $a\left( t\right) $ is the scale factor, $H=\dot{a}/a$ is the Hubble
function, and an overdot denotes differentiation with respect to cosmic time 
$t$. We adopt the metric and curvature conventions of \cite{wael}. The
dimensionless constant $Q$ is of the order $Q\lesssim 1$ \cite{amendola_coupled, amendola_viable} and parameterises
the strength of the interaction between the scalar field and the fluid; when 
$Q=0$ the two components evolve independently.

Interaction terms of the form $-\rho\dot{\phi}$ as in \eqref{conss} were
studied in \cite{bico} for exponential potentials, see also \cite{bclm}.
Although there is an energy exchange between the fluid and the scalar field,
it is easy to see that the set, $\rho >0,$ is invariant under the flow of
Eqs. \eqref{ray}--\eqref{conss}. This trivial physical requirement is
not satisfied if one assumes arbitrary interaction terms, cf. \cite{miri3}.

For even values of $n$, the potential satisfies $V(\phi)\ge 0$. Assuming
that initially $H(t_0)>0$, the Raychaudhuri equation \eqref{ray} implies
that $H$ is a strictly decreasing function of $t$. Since $V(\phi)\ge 0$, the Friedmann
equation \eqref{frie} prevents $H$ from changing sign, and the Universe
remains ever expanding.
Standard arguments (see Proposition~4 in~\cite{miri2}) show that the
matter density and the scalar kinetic energy both decay asymptotically and
the scalar field rolls monotonically toward the asymptotic regime 
$\phi(t)\to +\infty$ as  $t\to+\infty$. 
For odd values of $n$, the potential possesses a global maximum at some $\phi_m>0$ and is unbounded from below as $\phi\to -\infty$. In this case, for initially expanding models with $\phi(t_0)<\phi_m$ and insufficient initial velocity, the scalar field fails to cross the maximum and the evolution is driven toward the negative region of the potential. Under these conditions, the Universe collapses in a finite-time singularity. We refer the reader to~\cite{gmt1,gmt2} for a complete analysis.

Finite equilibria of the system \eqref{frie}--\eqref{conss} correspond to
stationary points of the potential, satisfying $V^{\prime }(\phi )=0$. For
the hybrid potential $V(\phi )=V_{0}\phi ^{n}e^{-k\phi }$, the condition $V^{\prime }(\phi )=0$ occurs at $\phi _{m}=n/k$, which corresponds to a local
maximum of the potential. The Friedmann constraint then gives $H^{2}=V(\phi
_{m})/3$, yielding a de Sitter solution. This equilibrium is denoted by $\mathcal{D}$ in Table \ref{t:crit} below. As expected for an equilibrium at
a maximum of the potential, this de Sitter point is unstable. In addition,
for $n\geq 2$, the potential satisfies $V(0)=0$ and $V^{\prime }(0)=0$,
giving rise to an equilibrium at 
\(
\phi =0, \dot{\phi}=0, \rho =0, H=0,
\)
corresponding to Minkowski spacetime. As such, the Minkowski solution does
not participate in the viable cosmological histories of the Universe, see
Section \ref{sec:viable}. 

We introduce expansion-normalised variables, \cite{clw,wael}, 
\begin{equation}
x=\frac{\dot{\phi}}{\sqrt{6}H},~~y=\frac{V}{3H^{2}},~~\Omega=\frac{\rho}{%
3H^{2}},  \label{env1}
\end{equation}
along with the effective steepness of the potential $z=-V^{\prime}/V=k-n/\phi
$. We also define a new time variable $\tau=\ln a$. Note that in order to
ensure that variable $z$ can consistently appear among the
expansion-normalised variables, we work with $\phi$ in reduced Planck units,
i.e. $\phi\to\phi/M_{\text{P}}$.

The definition of the variable $z$ becomes singular whenever the potential
satisfies $V(\phi)=0$. For the potential~\eqref{pote}, this occurs at $\phi=0
$ for all $n>0$. Consequently, if the cosmological evolution drives the
scalar field toward $\phi=0$, the variable $z$ diverges and the dynamical
system~\eqref{sys}, as formulated in the chosen variables, ceases to be
regular. For this reason, the present analysis is restricted to the region $%
\phi \neq 0$. This should be understood as a limitation of the chosen phase-space variables rather than as a singularity of the underlying cosmological system.

A distinguished feature of the present system is the set of invariant planes $z=k$. Indeed, from the evolution equation for $z$, one sees that $z=k$ is preserved by the flow. This plane corresponds to the asymptotic regime in which the logarithmic slope of the potential approaches the constant value $k$, and it will play a central role in what follows. In particular, the physically relevant critical points associated with the matter era and the late-time accelerated phase lie on this plane, so that viable cosmological histories are determined by the dynamics on and near $z=k$.

With the introduced variables, the Friedmann equation (\ref{frie}) takes the
form 
\begin{equation}
\Omega=1- x^{2}-y .  \label{cons1}
\end{equation}
Equation \eqref{cons1} can be used to eliminate $\Omega$ from the evolution
equations. We end up with the three-dimensional dynamical system, 
\begin{align}
\frac{dx}{d\tau} & = \sqrt{\frac{3}{2}}\frac{4-3\gamma}{2}%
Q\left(1-x^2-y\right)-3x+\sqrt{\frac{3}{2}}y z +3x^3+\frac{3}{2}\gamma
x\left(1-x^2-y\right) ,  \nonumber \\
\frac{dy}{d\tau} & = y\left( -\sqrt{6}xz+6x^2+3\gamma \left(1-x^2-y\right)
\right) ,  \label{sys} \\
\frac{dz}{d\tau} & = \sqrt{6}x\frac{\left(k-z\right)^2}{n}.  \nonumber
\end{align}
We recall the remarkable property of the Einstein equations that if the
constraint \eqref{frie} or \eqref{cons1} is satisfied at some initial time,
then it remains true throughout the evolution. In our case, this can be
verified explicitly from the evolution equations after a direct but lengthy
calculation. This result means that the surface in the phase space defined
by Eq.~\eqref{cons1} is an invariant manifold of the system, that is,
trajectories starting on it remain on it for all $\tau$. The physical
condition, $\Omega\geq 0$, restricts the phase space to 
\begin{equation}
x^{2}+y \le 1.  \label{phaseconstraint}
\end{equation}

Note that the standard choice of expansion--normalised variables, $y=\sqrt{V/(3H^{2})}$ automatically restricts analysis to non-negative potentials.
Definition \eqref{env1} allows the treatment of both positive and negative $%
V(\phi)$, which is essential for models featuring transitions for potentials
that may change sign. The drawback of the choice of $y$ in \eqref{env1} is
that the phase space is unbounded. That is, as $y\to-\infty$ the condition %
\eqref{phaseconstraint} allows $|x|$ to grow without bound, corresponding
to regimes where the potential is large and negative.

Finally, the effective equation of state, 
\[
w_{\mathrm{eff}}\equiv-1-\frac{2\dot{H}}{3H^{2}}, 
\]
takes the form, 
\begin{equation}
w_{\mathrm{eff}}=\gamma-1+(2-\gamma)x^{2}-\gamma y.  \label{weff}
\end{equation}

The critical points of the system \eqref{sys} together with the constraint %
\eqref{cons1} are listed in Table \ref{t:crit}. 
\begin{table}[!h]
\caption{Critical points of the system \eqref{sys}.}
\label{t:crit}%
\begin{tabular}{lll}
\hline
Label & $(x,y,z)$ & $\Omega$ \\ \hline
$\mathcal{A}_{\pm}$ & $\left( \pm1,0,k\right) $ & $0$ \\ 
$\mathcal{B}$ & $\left( \frac{\sqrt{\frac{2}{3}}\frac{ 4-3\gamma}{2} Q}{%
\left( 2-\gamma\right) },0,k\right) $ & $1-\frac{2}{3}\frac{\left( \frac{
4-3\gamma}{2}Q\right) ^{2}}{ \left( 2-\gamma\right)^{2} }$ \\ 
$\mathcal{C}$ & $\left( \frac{k}{\sqrt{6}},1-\frac{k^2}{6},k \right) $ & $0$
\\ 
$\mathcal{D}$ & $\left( 0,1,0\right) $ & $0$ \\ 
$\mathcal{E}$ & $\left( \frac{-\sqrt{\frac{3}{2}}\gamma}{\frac{4-3\gamma}{2}%
Q-k}, \frac{\left(\frac{4-3\gamma}{2}Q\right)^2+\frac{3\gamma}{2}(2-\gamma)-%
\frac{4-3\gamma}{2}Qk}{\left( \frac{4-3\gamma}{2}Q-k \right)^2},k \right) $
& $\frac{k \left(k-\frac{1}{2} (4-3 \gamma ) Q\right)-3 \gamma }{\left(k-%
\frac{1}{2} (4-3 \gamma ) Q\right)^2} $ \\ \hline
\end{tabular}
\vspace*{-4pt}
\end{table}

The linear stability of the equilibria is determined by the eigenvalues of
the Jacobian, presented in Table~\ref{t:eigen}. The corresponding scale-factor evolution is given in Table~\ref{t:sf}. The detailed
derivation of these results follows standard techniques of dynamical
systems, see for example \cite{perko}. 
\begin{table}[!h]
\caption{Eigenvalues of the linearisation at each critical point.}
\label{t:eigen}%
\begin{tabular}{ll}
\hline
Label & Eigenvalues \\ \hline
$\mathcal{A}_{\pm}$ & $0,~6\mp \sqrt{6}k,~3(2-\gamma)\mp\sqrt{6}\frac{%
4-3\gamma}{2}Q $ \\ 
$\mathcal{B}$ & $0,~\frac{\left(\frac{4-3\gamma}{2}Q\right)^{2}-\frac{3}{2}%
(2-\gamma)^2}{2-\gamma },~\frac{3\gamma(2-\gamma)+2\frac{4-3\gamma}{2}Q\left(%
\frac{4-3\gamma}{2}Q-k\right)}{2-\gamma} $ \\ 
$\mathcal{C}$ & $0,~\frac{1}{2}\left(k^{2}-6\right),-3\gamma-k\left(\frac{%
4-3\gamma}{2}Q-k\right)$ \\ 
$\mathcal{D}$ & $-3\gamma,~-\frac{3n+\sqrt{3}\sqrt{n(4k^2+3n)}}{2n},~ \frac{%
-3n+\sqrt{3}\sqrt{n(4k^2+3n)}}{2n} $ \\ 
$\mathcal{E}$ & $0, ~ \frac{\sigma+\sqrt{\sigma^2-4\delta }}{4\left(k- \frac{%
4-3\gamma}{2}Q\right)}, ~ \frac{\sigma-\sqrt{\sigma^2-4\delta }}{4\left(k- 
\frac{4-3\gamma}{2}Q\right)} $ \\ \hline
\end{tabular}
\vspace*{-4pt} \smallskip {\footnotesize \newline
where $\sigma=3 (\gamma -2) k+3 (4-3 \gamma ) Q$ and \newline
$\delta=6 \left(3 \gamma +k \left(\frac{1}{2} (4-3 \gamma )
Q-k\right)\right) \left(3 (\gamma -2) \gamma +(4-3 \gamma ) Q \left(k-\frac{1%
}{2} (4-3 \gamma) Q\right)\right)$ }
\end{table}

\begin{table}[!h]
\caption{Scale-factor evolution and effective equation of state
at each critical point.}
\label{t:sf}%
\begin{tabular}{lll}
\hline
Label & $a(t) $ & $w_\text{eff}$ \\ \hline
$\mathcal{A}_{\pm}$ & $t^{1/3}$ & $1$ \\ 
$\mathcal{B}$ & $t^{2(2-\gamma)/(3\gamma(2-\gamma)+2\left(\frac{4-3\gamma}{2}%
Q\right)^2)} $ & $-1+\gamma+\frac{2\left(\frac{4-3\gamma}{2}Q\right)^2}{%
3(2-\gamma)} $ \\ 
$\mathcal{C}$ & $t^{2/k^{2}}$ & $\frac{1}{3}\left(k^2-3\right)$ \\ 
$\mathcal{D}$ & $e^{t}$ & $-1$ \\ 
$\mathcal{E}$ & $t^{2(k-((4-3\gamma)/2)Q)/(3\gamma k)} $ & $\frac{(\gamma-1)
k+\frac{4-3\gamma}{2}Q}{k-\frac{4-3\gamma}{2}Q} $ \\ \hline
\end{tabular}
\vspace*{-4pt}
\end{table}

\section{Viable cosmology}

\label{sec:viable} The following analysis is based on the qualitative description of the evolution via a dynamical system approach. 
To assess the cosmological relevance of the equilibrium points, we impose the following qualitative requirements. A viable trajectory should pass near a transient matter-dominated or radiation-dominated solution, represented by a saddle point with the appropriate scale-factor behaviour, and then approach a late-time accelerated state. The analysis below shows that, among the admissible equilibria, this scenario is realised only for trajectories connecting the matter point $\mathcal{B}$ to the accelerated point $\mathcal{C}$. In order for a critical point to adequately describe the
matter-dominated epoch, it must satisfy three essential conditions:
\begin{itemize}
\item[(i)] \label{m1} Positive matter density, $\Omega > 0$; 
\item[(ii)] Correct scale factor evolution, $a \propto t^{2/3}$, that is
equivalent to $w_{\mathrm{eff}} \approx 0$; 
\item[(iii)] Transient nature, i.e. the point must be a saddle in the
dynamical systems sense.
\end{itemize}
Similarly, a viable late-time attractor must fulfill:
\begin{itemize}
\item[(iv)] Accelerated expansion, $w_{\mathrm{eff}} < -1/3$; 
\item[(v)] Stability, i.e. the point must be stable (sink).
\end{itemize}

We systematically explore the parameter space $(k, \gamma, Q)$ to identify
regions where both conditions are simultaneously satisfied, that is, a
transient matter era coexists with an accelerated future attractor.

\subsection*{Remarks on individual points}

\paragraph*{$\mathcal{A}_{\pm}, \mathcal{B}$---Zero-potential family $y=0$.}

These points correspond to regimes where the potential energy is negligible
compared to kinetic and matter components. They are determined by the
exponential tail of the potential, $z=k$ and describe kinetic-dominated or
scaling solutions. Points $\mathcal{A}_{\pm}=(\pm1,\,0,\,k) $ correspond to
purely kinetic cosmological states, lying on the boundary $x^{2}+y=1$ and
requiring $z=k$ so that $dz/d\tau=0$. In both cases, $\Omega=0$ and the
effective equation of state is $w_{\mathrm{eff}}=1$, indicating a stiff
fluid dominated by the scalar kinetic term. Hence, these points cannot
represent a matter-dominated phase nor an accelerated attractor. 

Point $\mathcal{B}$ corresponds to a scaling solution in which both the scalar
field and the matter sector contribute to the total energy density. It
enters the physically admissible phase space, $\Omega\geq0$, when 
\begin{equation}
Q \leq \sqrt{6}\,\frac{2-\gamma}{|\,4-3\gamma\,|},  \label{phaseB}
\end{equation}
for $\gamma\neq 4/3$ and is always allowed for $\gamma=4/3$. For typical
fluids with $1\leq \gamma < 4/3$, this condition is satisfied for all $%
Q\lesssim 1$. The condition for acceleration, $w_{\text{eff}}<-1/3$, is satisfied provided $(4-3\gamma)Q<\sqrt{2(2-\gamma)(2-3\gamma)}$ and $\gamma<2/3$; thus, point $\mathcal{B}$ cannot correspond to an accelerated attractor for any physically interesting values of $\gamma$.  Therefore point $\mathcal{%
B}$ remains as a candidate only for the matter epoch. Matter conditions (i)-(iii) are satisfied for 
\begin{equation}
Q=\frac{\sqrt{6\left( 2-\gamma\right) \left( 1-\gamma\right) }}{4-3\gamma}%
,~\gamma\leq1,~k<\sqrt{\frac{3}{2}}\sqrt{\frac{2-\gamma}{1-\gamma}}.
\label{Bmp}
\end{equation}

\paragraph*{$\mathcal{C}$---Scalar-field dominated point.}

Point $\mathcal{C }= \left(k/\sqrt6, 1-k^2/6, k\right)$
exists always for potentials taking negative values and only for $k\le \sqrt{%
6}$ when the potential is non negative. It can only be considered as a late
attractor since it does not satisfy the conditions for a matter point. Point 
$\mathcal{C}$ is accelerated and has non-positive eigenvalues whenever 
\begin{equation}
\left( 4-3 \gamma\right) Q> \frac{2\left( k^{2} -3 \gamma\right) }{k}~ \text{and} ~k<\sqrt{2}.   \label{pointC}
\end{equation}

Therefore, viable cosmological histories may occur only in the overlap of parameter values for which $\mathcal{B}$ represents a transient matter- or radiation-like saddle and $\mathcal{C}$ corresponds to an accelerated solution with two negative eigenvalues. 
Since one eigenvalue of the linearisation at $\mathcal{C}$ vanishes, linear theory is inconclusive. Restricting the flow to the local centre manifold (see Appendix \ref{sec:centre}) yields $du/d\tau = (k/n)u^2+(1/n)u^3+\mathcal{O}(u^4)$, where $u=z-k$, so the equilibrium is attracting for $u<0$ and repelling for $u>0$. In terms of the original variables, this corresponds to attraction on the $\phi>0$ branch and repulsion on the $\phi<0$ branch.

\paragraph*{$\mathcal{D}$---de Sitter}

Pure-potential point $\mathcal{D}=(0,1,0)$ although represents exponential acceleration with $a(t)\propto e^t$ is unstable (saddle point) and cannot
serve as a late-time attractor. Also, it does not satisfy the matter
condition $\Omega>0$ to be considered as a matter point.

\paragraph*{$\mathcal{E}$---Scaling solution}

Point $\mathcal{E}$ represents a scaling configuration where the scalar
field and matter coexist with fixed relative energy densities. It enters the
phase space when 
\begin{equation}
\left( 4-3\gamma\right) Q\leq\frac{2\left( k^{2}-3\gamma\right)}{k},
\label{phaseE1}
\end{equation}
We note that \eqref{phaseE1} contradicts \eqref{pointC}, therefore whenever
point $\mathcal{E}$ enters the phase space, $\mathcal{C}$ cannot be used for
the accelerated attractor phase. Therefore an acceptable trajectory cannot pass near point $\mathcal{E}$ to point $\mathcal{C}$ to represent the matter and accelerated epochs respectively. Point $\mathcal{E}$ cannot be used for
the accelerated epoch either, since it has to satisfy conditions (iv) and
(v) in Section~\ref{sec:viable}. These conditions imply 
\begin{equation}
\left( 4-3\gamma\right) Q<\left( 2-3\gamma\right) k~\text{ and }%
~\left(4-3\gamma\right) ^{2}Q^{2}-2k\left( 4-3\gamma\right)
Q+6\gamma\left(2-\gamma\right) >0.  \label{Eac}
\end{equation}
If point $\mathcal{E}$ is considered as an accelerated late attractor
satisfying the above conditions, then the only remaining candidate for the
matter epoch is point $\mathcal{B}$. However, conditions \eqref{Bmp} and %
\eqref{Eac} are incompatible for all the physically relevant values of $%
\gamma\geq 2/3$. 

In summary, point $\mathcal{E}$ cannot participate in a viable cosmological history, because the parameter conditions under which it can represent a transient matter-like epoch or a late-time accelerated phase are incompatible with those required for an other relevant point to be involved.

From the above analysis, we conclude that the only viable and physically
relevant trajectory is the one passing near the matter point $\mathcal{B}$
and being attracted to the late accelerated point $\mathcal{C}$, satisfying the
corresponding conditions \eqref{Bmp} and \eqref{pointC}.

Below we summarise the results for the physically relevant values $\gamma=1$
(dust), $\gamma=4/3$ (radiation) and $\gamma=2/3$ (ordinary matter
marginally satisfying the strong energy condition).

\paragraph{Dust ($\protect\gamma=1$).}
For dust, $\gamma=1$, the relevant critical points are collected in Table~\ref{t:g=1}.
\begin{table}[!h]
\caption{Critical Points for the case of dust ($\protect\gamma=1$) with
coupling constant $Q \lesssim 1$ and parameter $k<\protect\sqrt{2}$.}
\label{t:g=1}%
\begin{tabular}{llll}
\hline
Label $(x,y,z)$ & $\Omega$ & Stability & $a(t)$ \\ \hline
$\mathcal{A}_{\pm}$ $\left( \pm 1,0,k\right) $ & $0$ & Unstable & $t^{1/3}$
\\ 
$\mathcal{B}$ $\left( \frac{ Q}{\sqrt{6}},0,k\right) $ & $1-\frac{Q^{2}}{6}$
& Saddle & $t^{4/ \left( 6 + Q^{2} \right) }$ \\ 
$\mathcal{C}$ $\left( \frac{k}{\sqrt{6}},1-\frac{k^{2}}{6},k\right) $ & $0$
& Conditionally saddle & $t^{2/ k^{2}}$ \\ 
$\mathcal{D}$ $\left( 0,1,0 \right) $ & $0$ & Saddle & $e^{t} $ \\ 
$\mathcal{E}$ $\left( \frac{\sqrt{6}}{2k-Q},\frac{Q^2-2kQ+6}{(Q-2k)^2},k
\right) $ & $\frac{-12+4k^2-2kQ}{(Q-2k)^2}$ & Saddle & $t^{2 /3-Q/(3k)} $ \\ 
\hline
\end{tabular}
\vspace*{-4pt}
\end{table}

In this case, the requirement of a standard matter era forces the coupling to vanish, $Q=0$. This is one of the main physical conclusions of the analysis: in the present model, a nonzero coupling is incompatible with the conventional dust-dominated phase $a(t)\propto t^{2/3}$. On the invariant plane $z=k$, the matter point $\mathcal{B}$ is a saddle, while the accelerated point $\mathcal{C}$ is conditionally attracting: trajectories with $\phi>0$ approach $\mathcal{C}$, whereas trajectories with $\phi<0$ are repelled from it. A viable cosmological history is therefore represented by a trajectory that passes near $\mathcal{B}$ and then approaches $\mathcal{C}$. By contrast, trajectories on the $\phi<0$ branch do not settle at the accelerated point and instead evolve toward collapse in finite time when $n$ is odd.

\begin{figure}[!h]
\centering
\includegraphics[width=2.5in]{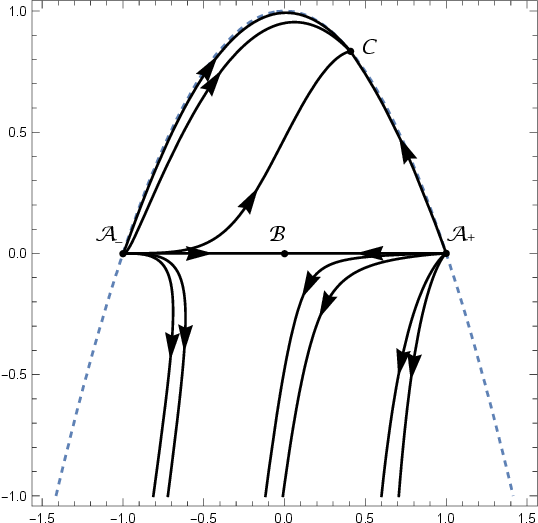}
\caption{Phase portrait on the invariant plane $z=k$ (shown here for $k=1$) in the dust case $\gamma=1$ with $Q=0$. The dashed parabola marks the boundary of the phase space. The saddle point $\mathcal{B}$ represents the transient matter era, while $\mathcal{C}$ corresponds to the accelerated phase. Trajectories passing near $\mathcal{B}$ and approaching $\mathcal{C}$ describe viable cosmological histories. Divergent trajectories correspond to the $\phi<0$ branch and, for odd $n$, lead to collapse in finite time.}
\label{fig:ps1}
\end{figure}

The phase portrait illustrated in Fig. \ref{fig:ps1}, highlights the distinction between non-negative and sign-changing potentials. 
For even values of $n$, where $V(\phi)\geq 0$, the phase space is bounded and trajectories on the expanding branch are driven toward the accelerated solution $\mathcal{C}$. 
In contrast, for odd values of $n$, where the potential becomes negative for $\phi<0$, the accelerated point is only conditionally stable: trajectories with $\phi>0$ approach $\mathcal{C}$, while those with $\phi<0$ are driven away, leading the Universe to collapse in a finite time.

\paragraph{Radiation era ($\protect\gamma = 4/3$).}
For $\gamma=4/3$, the interaction terms in \eqref{ray} and \eqref{ems} vanish identically, and the system admits a transient radiation-dominated phase with the correct behaviour $a(t)\propto t^{1/2}$. When $k<\sqrt{2}$, point $\mathcal{C}$ yields accelerated expansion. However, this acceleration is again only conditional: on the invariant plane $z=k$, point $\mathcal{C}$ is attracting for trajectories with $\phi>0$, whereas off the plane it behaves as a saddle and repels the $\phi<0$ branch. This branch dependence is precisely the one identified by the centre-manifold reduction discussed above: the late-time accelerated state is attracting only on the side corresponding to $z<k$ (equivalently $\phi>0$). Thus the model admits a viable radiation-to-acceleration history, but not a global late-time attractor.

\paragraph{Case $\protect\gamma = 2/3$.}
For $\gamma = 2/3$, corresponding to matter marginally satisfying the strong energy condition, the correct matter-era scaling $a(t)\propto t^{2/3}$  fixes the coupling to $Q=\sqrt{2/3}$. In this case the point $\mathcal{B}=(1/2,0,k)$ is a saddle with $\Omega=3/4$, and therefore represents a transient matter-dominated phase. For $k<\sqrt{2}$, point $\mathcal{C}$ again gives accelerated expansion and, as in the previous cases, acts as an attractor only for the branch $\phi>0$. The resulting cosmological history is therefore viable, but only conditionally so.

\bigskip

\section{Conclusion}
\label{sec:concl}

The present analysis shows that the hybrid potential $V(\phi) = V_{0}\,\phi^{n}e^{-k\phi}$ supports viable cosmological histories only under restricted conditions. The central geometric feature are the invariant planes $z=k$, which contain both the transient matter point $\mathcal{B}$ and the accelerated point $\mathcal{C}$. In all viable cases, cosmological evolution proceeds along or near these planes from a saddle matter or radiation phase toward late-time acceleration.

A physically significant restriction concerns the coupling $Q$. For dust, $\gamma=1$, a standard matter era requires $Q=0$, so the conventional dust epoch is incompatible with nonzero coupling in this model. For $\gamma=2/3$, the correct matter-era scaling fixes $Q=\sqrt{2/3}$, while for radiation, $\gamma=4/3$, the interaction term vanishes identically. In all three cases, accelerated expansion requires $k<\sqrt{2}$.


\paragraph*{Significance of the qualitative properties of the potential.}
At this point we note a striking similarity of the qualitative behaviour of
the solutions of our dynamical system \eqref{sys} and the system corresponding
to a scalar field with a double exponential potential, $V\left(  \phi\right)
=V_{1}e^{-\alpha\phi}+V_{2}e^{-\beta\phi},$ \cite{tzmi}. Firstly, equilibria
denoted by $\mathcal{B}$ in both systems, with analogous coordinates,
represent a transient matter dominated epoch with the same scale factor
dependence, $a\left(  t\right)  \propto t^{2/3}$ for $Q$ close to zero.
Likewise, equilibria denoted by $\mathcal{C}$ in both systems, with analogous
coordinates represent a future accelerated epoch with the same scale factor
dependence for $Q$ close to zero. Similarly, equilibria $\mathcal{A}_{\pm}$
and $\mathcal{D}$ share the same properties in both systems. Secondly in both
systems, trajectories passing near the matter point $\mathcal{B},$
asymptotically approach the future accelerating attractor $\mathcal{C}$.

This comparison supported by the conclusions on general negative potentials in
\cite{gmt1,gmt2}, allow us to conjecture that the qualitative cosmological
dynamics is largely insensitive to the precise functional form of the
scalar-field potential. Instead, it is the global qualitative properties of
the potential, such as the existence of a local maximum, its asymptotic
behaviour at large field values, and whether it attains negative values, that
determine the structure of the phase space and the nature of viable
cosmological trajectories. In particular, these features control the presence
of unstable de-Sitter solutions, transient matter phases, accelerated
attractors at late times, or eventual recollapse. Once the essential
qualitative features of the potential are fixed, the resulting cosmic
evolution is probably largely constrained, and the conclusions drawn in this
work may be extend beyond the specific potential considered here. 

\vskip6pt

\acknowledgments{The authors would like to thank the Guest Editors, Spiros Cotsakis and
Alexander Yefremov, for the invitation to contribute to the Special Issue of
the broader theme on Dynamical Systems and Stability in Cosmology and
Gravitation.}

\bigskip \appendix
\numberwithin{equation}{section} \renewcommand{\theequation}{\thesection.%
\arabic{equation}}

\bigskip \numberwithin{equation}{section} \renewcommand{\theequation}{%
\thesection.\arabic{equation}}

\section{Centre manifold reduction at the accelerated point}

\label{sec:centre}

In this appendix we determine the local dynamics near the non--hyperbolic
accelerated critical point $\mathcal{C}$ by means of centre manifold theory 
\cite{guho}. This analysis is required because the linearisation at $%
\mathcal{C}$ possesses a zero eigenvalue, rendering linear stability
arguments inconclusive.

We consider the autonomous system~\eqref{sys} and the equilibrium 
\[
\mathcal{C}=\left(\frac{k}{\sqrt{6}},\,1-\frac{k^2}{6},\,k\right). 
\]
Translating the origin, we define 
\[
X := x-\frac{k}{\sqrt{6}}, \qquad Y := y-1+\frac{k^2}{6}, \qquad Z := z-k, 
\]
and the system can be written in the form 
\begin{equation}
\frac{d}{d\tau} \begin{pmatrix} X \\ Y \\ Z \end{pmatrix} = A %
\begin{pmatrix} X \\ Y \\ Z \end{pmatrix} + \mathbf{F}(X,Y,Z),
\label{eq:cm_full}
\end{equation}
where $A$ is the Jacobian matrix at the origin and $\mathbf{F}$ contains
nonlinear terms of quadratic and higher order. The eigenvalues of $A$ are 
\[
\lambda_1 = 0, \qquad \lambda_2 = \frac{1}{2}(k^2-6), \qquad \lambda_3 =
-3\gamma + k^2 - \frac{k}{2}(4-3\gamma)Q. 
\]
Under the condition~\eqref{pointC}, one has $\lambda_2<0$ and $\lambda_3<0$.
Hence the equilibrium admits a one--dimensional centre subspace $E^c$
associated with $\lambda_1=0$, and a two--dimensional stable subspace $E^s$.

We perform a linear change of variables $\mathbf{U}=T^{-1}\,\mathbf{X}$ with 
$\mathbf{X}=(X,Y,Z)^\top,~\mathbf{U}=(u,v,w)^\top$, where $T$ is the matrix
whose columns are the eigenvectors of $A$. In the new coordinates the linear
part is block-diagonal with a $1\times 1$ center block, 
\[
T^{-1}AT= \begin{pmatrix}
0 & 0 \\
0 & P
\end{pmatrix}, 
\]
where $P$ is a $2\times 2$ matrix with eigenvalues $\lambda_2,\lambda_3$.

With this change of variables, the system can be brought in the standard form

\begin{align}
\frac{du}{d\tau} &= F(u,v,w),  \label{eq:cm_u} \\
\frac{d}{d\tau} \begin{pmatrix} v \\ w \end{pmatrix} &= P \begin{pmatrix} v
\\ w \end{pmatrix} + \mathbf{G}(u,v,w),  \label{eq:cm_vw}
\end{align}
where $u=Z$ characterises the centre subspace, $(v,w)$ span the stable
subspace, $P$ has eigenvalues $\lambda_2,\lambda_3$, and $F, \mathbf{G}$ are the non-linear parts of the vector field with
\[
F(0,0,0)=0, \qquad DF(0,0,0)=0, \qquad \mathbf{G}(0,0,0)=\mathbf{0}, \qquad D%
\mathbf{G}(0,0,0)=0. 
\]

The centre manifold theorem guarantees \cite{guho} the existence of a
locally invariant one--dimensional manifold of the form 
\[
W^{\text{c}} = \{ (u,v,w): v=h_1(u),\; w=h_2(u),\; |u|<\delta \}, 
\]
with 
\[
h_1(0)=h_2(0)=0, \qquad h_1^{\prime }(0)=h_2^{\prime }(0)=0. 
\]
The function $\mathbf{h}(u)=(h_1(u),h_2(u))^\top$ satisfies the equation, (see
\cite{perko}) 
\begin{equation}
D\mathbf{h}(u)\,F(u,\mathbf{h}(u)) = P\,\mathbf{h}(u) + \mathbf{G}(u,\mathbf{%
h}(u)).  \label{eq:invariance}
\end{equation}
Since $\mathbf{h}(0)=\mathbf{0}$ and $D\mathbf{h}(0)=\mathbf{0}$, we seek a
Taylor expansion starting with second order terms, 
\[
h_1(u)=a_2 u^2+a_3 u^3+\mathcal{O}(u^4), \qquad h_2(u)=b_2 u^2+b_3 u^3+%
\mathcal{O}(u^4). 
\]
Substituting these expansions into~\eqref{eq:invariance} and matching
coefficients of like powers of $u$ yields the coefficients $a_i$, $b_i$.
Their explicit expressions are lengthy and are given below for completeness, 
\begin{align}
a_2&= \frac{2 k^2}{18 n-3 k^2 n}-\frac{(\gamma -1) (k^2-6)}{6 \left(6 \gamma
-k^2+(4-3 \gamma ) k Q-6\right)}-\frac{1}{6}, \\
a_3&= \frac{2}{3 n^2} \left(\frac{(\gamma -1) k n}{-6 \gamma +k^2-(4-3
\gamma ) k Q+6}-\frac{4 k (k^2-3 n)}{(k^2-6)^2}\right), \\
b_2&= \frac{(\gamma -1) (k^2-6)}{6 \left(6 \gamma -k^2+(4-3 \gamma ) k
Q-6\right)}, \\
b_3&= \frac{2 (\gamma -1) k}{3 n \left(6 \gamma -k^2+(4-3 \gamma ) k
Q-6\right)}.
\end{align}

The dynamics on the centre manifold is obtained by substituting $%
(v,w)=(h_1(u),h_2(u))$ into~\eqref{eq:cm_u}, yielding 
\begin{equation}
\frac{du}{d\tau} = \frac{k}{n}u^2 +\frac{1}{n}u^3 +\mathcal{O}(u^4).
\label{eq:centrem}
\end{equation}
The leading nonzero term is quadratic and positive for $k>0$ and $n>0$.
Consequently, 
\[
\frac{du}{d\tau}\sim \frac{k}{n}u^2, \qquad u\to 0. 
\]
It follows that $u=0$ is a saddle on the centre manifold; trajectories with $%
u<0$ are attracted toward the equilibrium, while trajectories with $u>0$ are
repelled. Translating back to the original variables, this implies that the
accelerated point $\mathcal{C}$ is stable for $z<k$, equivalently $\phi>0$,
and unstable for $z>k$ ($\phi<0$), establishing the conditional nature of
late--time acceleration discussed in the main text.

\bibliographystyle{unsrt}
\bibliography{powerexpo}
\end{document}